\documentclass[british,american,prl,final,showpacs,superscriptaddress,showkeys,lengthcheck]{revtex4}
\usepackage[T1]{fontenc}
\usepackage[latin9]{inputenc}
\usepackage{color}
\usepackage{babel}
\usepackage{textcomp}
\usepackage{amsmath}
\usepackage{amssymb}
\usepackage{graphicx}
\usepackage{esint}
\usepackage[unicode=true,pdfusetitle,
 bookmarks=true,bookmarksnumbered=false,bookmarksopen=false,
 breaklinks=false,pdfborder={0 0 1},backref=false,colorlinks=true]
 {hyperref}

\makeatletter
\@ifundefined{textcolor}{}
{%
 \definecolor{BLACK}{gray}{0}
 \definecolor{WHITE}{gray}{1}
 \definecolor{RED}{rgb}{1,0,0}
 \definecolor{GREEN}{rgb}{0,1,0}
 \definecolor{BLUE}{rgb}{0,0,1}
 \definecolor{CYAN}{cmyk}{1,0,0,0}
 \definecolor{MAGENTA}{cmyk}{0,1,0,0}
 \definecolor{YELLOW}{cmyk}{0,0,1,0}
}

\@ifundefined{definecolor}{color}{}
\@ifundefined{definecolor}{\@ifundefined{definecolor}
 {\usepackage{color}}{}
}{}

\@ifundefined{definecolor}{\@ifundefined{definecolor}
 {\@ifundefined{definecolor}
 {\usepackage{color}}{}
}{}
}{}
\usepackage{babel}
\usepackage{longtable}

\@ifundefined{textcolor}{}{%
 \definecolor{BLACK}{gray}{0}
 \definecolor{WHITE}{gray}{1}
 \definecolor{RED}{rgb}{1,0,0}
 \definecolor{GREEN}{rgb}{0,1,0}
 \definecolor{BLUE}{rgb}{0,0,1}
 \definecolor{CYAN}{cmyk}{1,0,0,0}
 \definecolor{MAGENTA}{cmyk}{0,1,0,0}
 \definecolor{YELLOW}{cmyk}{0,0,1,0}
 }

\usepackage{epsfig}

\usepackage{amsfonts}\usepackage{float}

\usepackage{babel}

\makeatother

\begin{document}

\title{Criticality of environmental information obtainable by dynamically
controlled quantum probes}

\author{Analia Zwick, Gonzalo A. Álvarez and Gershon Kurizki}

\affiliation{Weizmann Institute of Science, Rehovot 76100, Israel}
\begin{abstract}
A universal approach to decoherence control combined with quantum
estimation theory reveals a critical behavior, akin to a phase transition,
of the information obtainable by a qubit probe concerning the memory
time of environmental fluctuations. This criticality emerges only
when the probe is subject to dynamical control. It gives rise to a
sharp transition between two dynamical phases characterized by either
a short or long memory time compared to the probing time. This phase-transition
of the environmental information is a fundamental feature that facilitates
the attainment of the highest estimation precision of the environment
memory-time and the characterization of probe dynamics.
\end{abstract}


\maketitle
\medskip{}

\noindent A simple quantum-probe, such as a qubit, is capable of extracting
information on the environment dynamics and its space-time fluctuations
through the spectrum of the dephasing noise the probe is subject to
\cite{Mittermaier2006,Taylor2008,Balasubramanian2008,almog_direct_2011,Alvarez2011,Bylander2011,Smith2012,Alvarez2013,Kucsko2013,Neumann2013,Toyli2013,Steinert2013,Grinolds2014,Sushkov2014}.
This information is the subject of an emerging field of research dubbed
environmental quantum-noise spectroscopy \cite{Alvarez2011,Bylander2011}.
Its most straightforward implementation is by monitoring the free-induction
decay of an initially-prepared qubit-probe coherence and inferring
the dephasing characteristics from this decay \cite{Abragam1961a,Slichter1990}.
A more promising option is to exert a control (driving) field, whether
pulsed or continuous-wave (CW), on the qubit-probe and study its dephasing
as a function of the control-field characteristics \cite{almog_direct_2011,Alvarez2011,Bylander2011}.
Pulsed control of qubit dephasing is commonly described by dynamical
decoupling \cite{vio98,vio99,Zanardi1999,Khodjasteh2005,Uhrig2007}.
However, for the purpose of environment-noise spectroscopy it is useful
to resort to the universal formula for the rate of decoherence under
dynamical control \cite{kofman_unified_2004,Kofman_Theory_2005,kofman_universal_2001,clausen_bath-optimized_2010},
which is at the heart of the unified theory of dynamically-controlled
open quantum systems \cite{gordon_universal_2007,gordon_optimal_2008,Zwick_ChapGK_2015}.
This universal formula allows the design of control fields or pulse
sequences that through the choice of a spectral filter function are
optimally tailored to the specific environment-noise spectrum and
the task at hand \cite{clausen_bath-optimized_2010,clausen_task-optimized_2012}:
decoherence control \cite{kofman_unified_2004,Kofman_Theory_2005,gordon_universal_2007,gordon_optimal_2008,gordon_Preventing_2006,gordon_Dynamical_2008,Uys2009,KLV10,Ajoy2011},
state-transfer \cite{Zwick_Optimized_2014,escher_optimized_2011}
or storage \cite{petrosyan_reversible_2009,bensky_optimizing_2012,escher_optimized_2011}
in a fluctuating environment. Here, the filter function will be optimally
adapted to the task of probing the environment-noise spectrum by a
qubit \cite{Zwick_NatCom_2015}.

Among environment-noise parameters whose estimation is of practical
interest in physics, chemistry and biology, the \emph{memory or correlation
time} is particularly helpful \cite{almog_direct_2011,Alvarez2011,Bylander2011,Smith2012,Alvarez2013,Alvarez2013_JMR,Zwick_NatCom_2015,Shemesh2015,Feintuch2004,Suter1985,'Alvarez2015,Alvarez2013a,Haikka2012,Gessner2014,Smirne2013,Laine2012,Liu2013,Wissmann2014}.
On a fundamental level, memory effects of the environment are associated
with the concept of non-Markovianity, whose definition is an outstanding
issue \cite{Rivas2014,Breuer2015}. 

For the purpose of characterizing the memory effects of an environment
that interacts with a qubit probe, we here put forward an approach
based on the aforementioned universal formula for decoherence control
\cite{kofman_unified_2004,Kofman_Theory_2005,kofman_universal_2001,clausen_bath-optimized_2010,clausen_task-optimized_2012,gordon_universal_2007,gordon_optimal_2008,Zwick_ChapGK_2015}
combined with quantum estimation theory \cite{Caves_1994_fisher,Paris2009_QUANTUM-ESTIMATION,escher2011general,Paris2014_Characterization-of-classical-Gaussian,Zwick_NatCom_2015}.
We show that the information (estimation-precision) concerning the
environment-noise fluctuation-spectrum obtained by this approach may
exhibit \emph{critical behavior} as a function of the memory-time
parameter. This critical behavior, akin to a phase transition, is
only revealed under dynamical control: it defines a \emph{sharp boundary}
between the short- and long-time regimes of the probe decoherence
corresponding to long- and short-memory of the environment respectively.
By contrast, free-induction decay of the probe coherence undergoes
the usual \emph{smooth transition} between the two-regimes, thus conforming
to the gradual change from non-Markovianity to Markovianity that has
been previously analyzed \cite{Rivas2014,Breuer2015}. The criticality
or phase-transition of the environmental information revealed here
is a fundamental feature that characterizes dynamical behavior with
practical implications on the attainment of the highest estimation
precision of the environment memory-time (see Discussion).

\paragraph*{Controlled qubit-probe as a sensor of the environmental fluctuations\emph{.--- }}

\begin{figure}
\includegraphics[width=1\columnwidth]{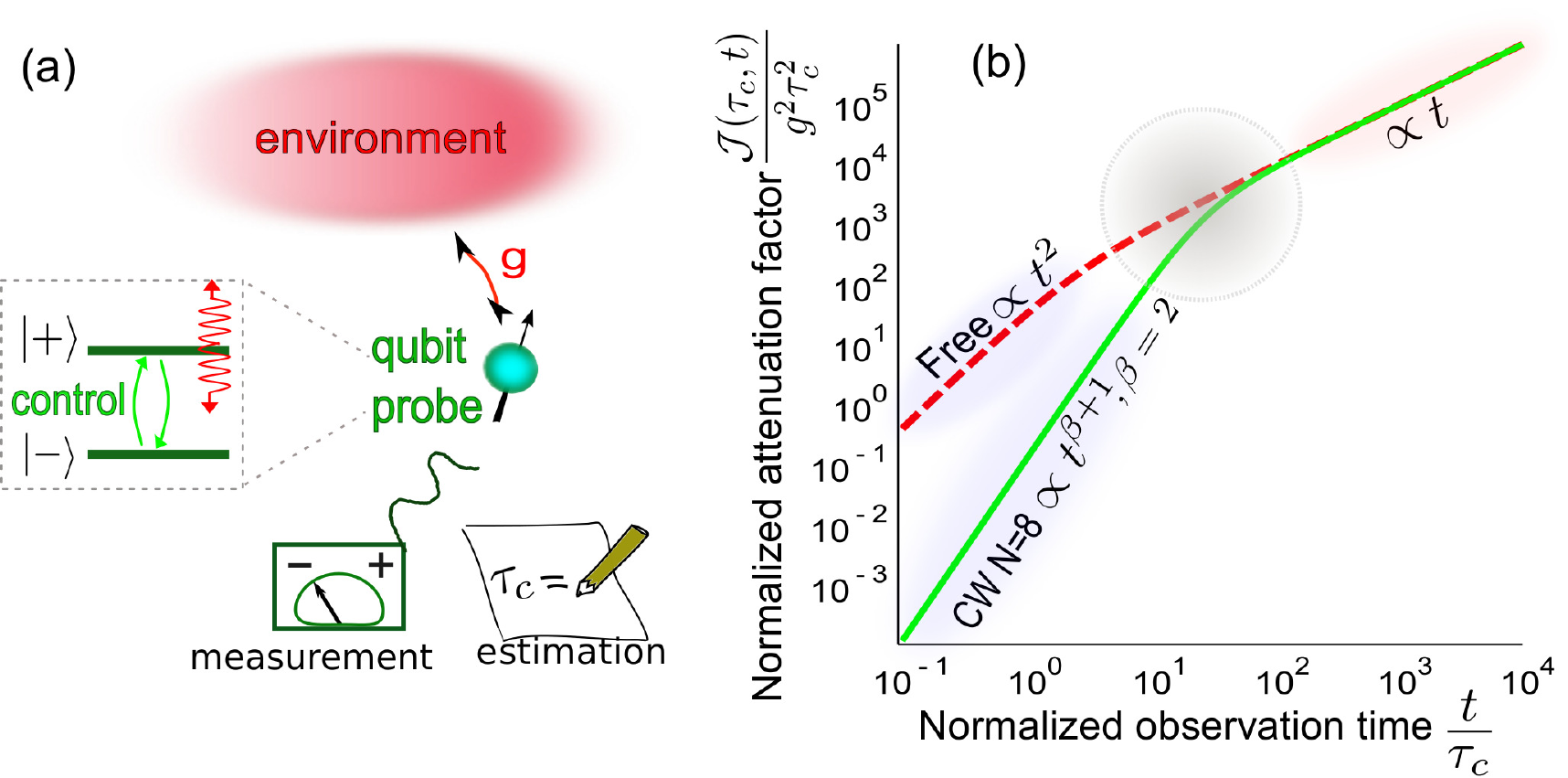}

\protect\caption{\label{fig1_scheme}(a) Estimation of the environmental noise fluctuations
by a qubit that probes the environment. A dynamically controlled qubit-probe
undergoes pure-dephasing by the environment due to the interaction
$H_{SB}=g\sigma_{z}B$, where $\sigma_{z}$ is the Pauli operator
for the probe and $B$ is the environment (bath) operator. The dephasing
is revealed by the attenuation (decay) factor $\mathcal{J}$ that
characterizes the optimal qubit-observable $\left\langle \sigma_{x}(\vec{x}_{B},t)\right\rangle =e^{-\mathcal{J}(\vec{x}_{B},t)}$,
$\vec{x}_{B}$ being the set (unknown) noise parameters, for an (optimal)
initial state - the symmetric superposition of the spin-up/-down states
in the basis $\sigma_{z}$, $\frac{1}{\sqrt{2}}(\left|\uparrow\right\rangle +\left|\downarrow\right\rangle )=\left|+\right\rangle $
\cite{Paris2014_Characterization-of-classical-Gaussian,Zwick_NatCom_2015,Zwick_PhysScrip_2015}.
Here we focus in estimating $x_{B}=\tau_{c}$. (b) Time dependence
of the normalized attenuation factor $\mathcal{J}$ of the qubit-state
probing an \textit{\emph{Ornstein-Uhlenbeck}} process ($\beta=2$)
for free evolution (dashed) compared to its counterpart under dynamical
control (solid). The latter time dependence exhibits a smooth transition
(marked by a circle) between two well-defined dynamical phases (regimes)
depending on the ratio $\frac{t}{\tau_{c}}$. }
\end{figure}

We consider a \emph{dynamically-controlled} qubit-probe experiencing
pure dephasing in the \emph{weak-coupling probe-environment regime}
(Fig. \ref{fig1_scheme}a), which is characterized by the attenuation
(decay) factor $\mathcal{J}(\vec{x}_{B},t)$ of the qubit coherence
\cite{Zwick_NatCom_2015} 
\begin{equation}
\left\langle \sigma_{x}(t)\right\rangle =\sigma_{x}(0)e^{-\mathcal{J}(\vec{x}_{B},t)},\label{eq:ox}
\end{equation}
where $\vec{x}_{B}$ are a set of parameters that describe the environment
and the attenuation factor obeys the universal formula \cite{kofman_unified_2004,Kofman_Theory_2005,kofman_universal_2001,clausen_bath-optimized_2010,clausen_task-optimized_2012,gordon_universal_2007,gordon_optimal_2008,Zwick_ChapGK_2015}
\begin{equation}
\mathcal{J}(\vec{x}_{B},t)=\int_{-\infty}^{\infty}d\omega F_{t}(\omega)G(\vec{x}_{B},\omega).\label{eq:Dephasing rate xB}
\end{equation}
Here $G(\vec{x}_{B},\omega)$ is the coupling spectrum (spectral density)
of the environment noise (the Fourier transform of its autocorrelation
function). Explicitly, $\vec{x}{}_{B}=\left[g,\tau_{c},\beta\right]$,
with $\tau_{c}$ as the correlation or memory time of the environment
noise, \emph{i.e.} the inverse width of its spectral density, $g$
as the \emph{effective} probe-environment coupling-strength, and $\beta$
as a power law exponent that defines the type of stochastic (noise)
process. The filter function $F_{t}(\omega)$ explicitly depends upon
the dynamical control of the probe during time $t$. The information
about the unknown environment parameters $\vec{x}_{B}$ is encoded
by the probabilities $p$ of finding the qubit-probe in the $\left|+\right\rangle $
(symmetric) or $\left|-\right\rangle $ (antisymmetric) superposition
of the qubit energy states when measuring $\sigma_{x}$. These probabilities
obey 
\begin{equation}
p_{\pm}(\vec{x}_{B},t)=\frac{1}{2}\left(1\pm e^{-\mathcal{J}(\vec{x}_{B},t)}\right).\label{eq:p+-}
\end{equation}

As a model to describe the memory time scales of the environment,
we consider \textit{\emph{a }}\textit{generalized Ornstein-Uhlenbeck
spectral density} \cite{Zwick_NatCom_2015}
\begin{equation}
G_{\beta}(\left[g,\tau_{c},\beta\right],\omega)=g^{2}\frac{\mathcal{A}_{\beta}\tau_{c}}{1+\omega^{\beta}\tau_{c}^{\beta}},\label{eq:power-law tail noise spectrum}
\end{equation}

\noindent where $\mathcal{A}_{\beta}$ is a normalization factor depending
on the power-law $\beta\ge2$. In fact, this model is the building
block of \emph{universal} lineshapes: it may characterize the memory
time of \emph{arbitrary bosonic environments}, if one assumes that
a chosen harmonic-oscillator mode constitutes an interface between
the qubit-probe and the modes of the environment \cite{kofman1994spontaneous}.
The combined spectrum of \emph{any environment} plus the interface
mode is then reshaped, or \textquotedbl{}filtered\textquotedbl{},
according to the chosen oscillator-mode frequency and its coupling
strength with the probe, invariably resulting in a\textit{ skewed-Lorentzian
lineshape} \cite{kofman1994spontaneous,PhysRevE.87.012140}. 

\begin{figure}[t]
\centering{}

\includegraphics[width=0.9\columnwidth]{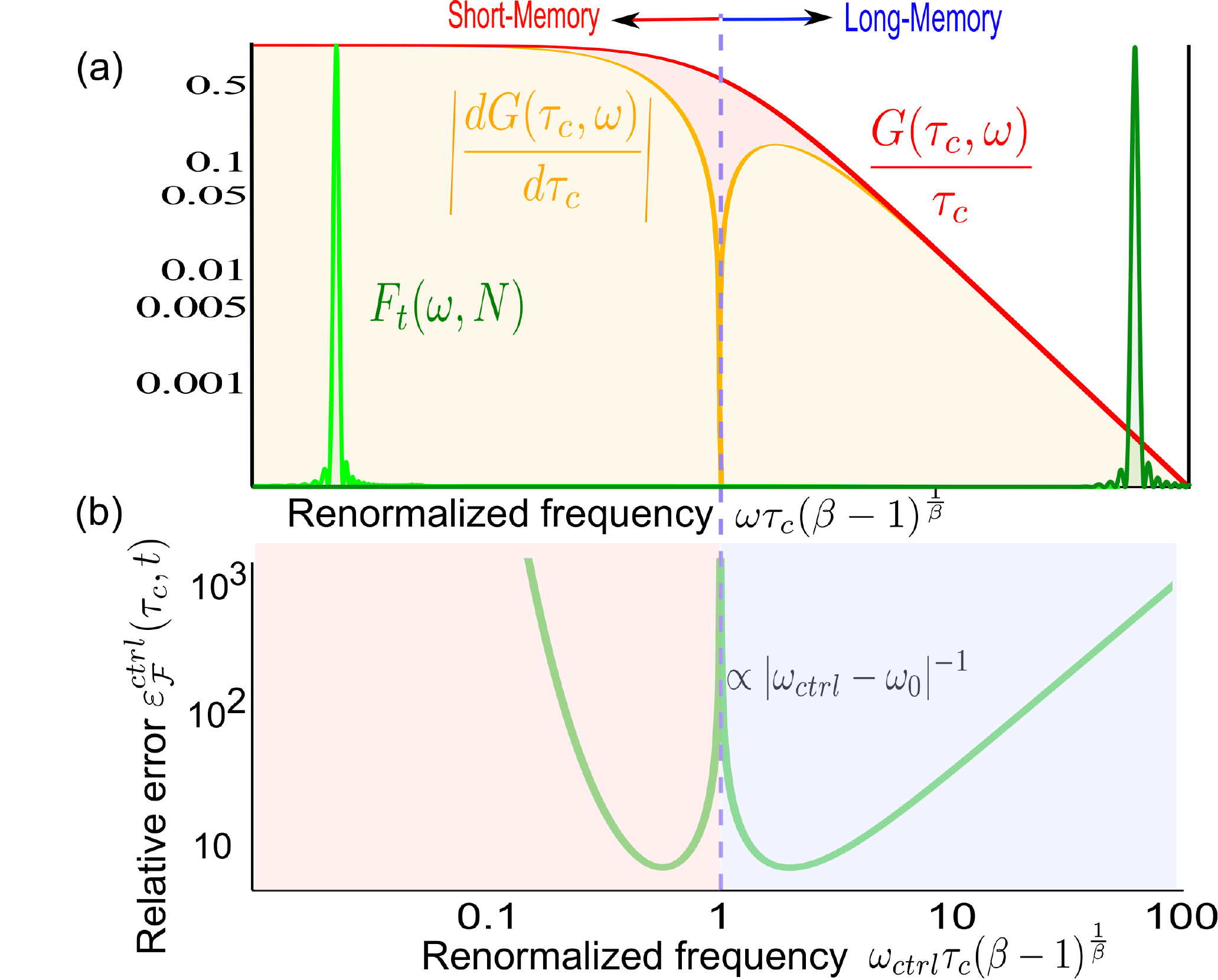}

\protect\caption{\label{fig:FtwGw} \textit{\emph{Criticality of the probe-extracted
information on the environmental correlation (memory) time $\tau_{c}$.}}
(a) Spectral density\textit{\emph{ for the Ornstein-Uhlenbeck process}}
(Lorentzian spectrum, $\beta=2$: red solid line). The derivative
of the spectrum $\left|\frac{dG}{d\tau_{c}}\right|$ exhibits a critical
behavior $\left|\frac{dG}{d\tau_{c}}\right|\propto\left|\omega-\omega_{0}\right|$
at $\omega_{0}=\tau_{c}^{-1}(\beta-1)^{-\frac{1}{\beta}}$. The two
dynamical regimes occur when the narrow filters probe frequency components
of $G(\omega)$ on both sides of the critical point. Two typical continuous
wave (CW-control) filter functions $F_{t}(\omega)$ (green, in linear
scale) scan the spectrum on both sides of the transition ($N=20$).
(b) The attainable relative error $\varepsilon_{\mathcal{F}}^{ctrl}(\tau_{c},t=\frac{\pi N}{\omega_{ctrl}})$
on $\tau_{c}$ by the qubit-probe under CW control ($\sqrt{2N}g\tau_{c}=1$,
$\beta=2$ and $\omega_{ctrl}=\frac{\pi N}{t}$). The divergence $\propto\left|\omega_{ctrl}-\omega_{0}\right|^{-1}$
at the critical point gives evidence of the critical behavior.}
\end{figure}

The power law regime $\propto\omega^{-\beta}$ of the spectral density,
obtained for $\omega\tau_{c}\gg1$, is the spectral range with the
strongest dependence on the frequency $\omega$, describing the short
time behavior of the probe-qubit dephasing. We define this limit as
the Long Memory (LM) regime\textit{. }In the opposite limit $\omega\tau_{c}\ll1$,
associated with long times, the spectral density becomes independent
of the frequency, and the probe coherence attenuation factor $\mathcal{J}(\vec{x}_{B},t)$
is then given by the Fermi Golden Rule. We dub this limit the Short
Memory (SM) regime.

\paragraph{\noindent Identifying the dynamical regimes' criticality by dynamically
controlled probes.---}

Under free induction decay, the LM and SM regimes are attained at
times $t\ll\tau_{c}$ and $t\gg\tau_{c}$, respectively. The respective
decay (attenuation) factors are $\mathcal{J}_{free}^{LM}\propto g^{2}t^{2}$
(independent of $\tau_{c}$) and $\mathcal{J}_{free}^{SM}\propto\! g^{2}\tau_{c}t$,
by considering $F_{t}^{free}(\omega)\!=\!\frac{t\text{\texttwosuperior}\mathrm{sinc^{2}(\frac{\omega t}{2})}}{2}$
\cite{Zwick_NatCom_2015}. The transition from the LM to the SM regime
is smooth (Fig. \ref{fig1_scheme}b) as the ratio $\frac{t}{\tau_{c}}$
is varied and does not depend on $g$. Invariably, $\frac{\partial\mathcal{J}_{free}}{\partial\tau_{c}}>0$,
\emph{without} sign change.

Consider now the change that may arise in the character of this transition
under dynamical control. An example is a decoupling control sequence
of $N\gg1$ equidistant $\pi$-pulses (known as CPMG) \cite{Carr1954,Meiboom1958,Slichter1990}.
The filter function $F_{t}(\omega)$ \cite{almog_direct_2011,kofman_unified_2004,Kofman_Theory_2005,kofman_universal_2001,clausen_bath-optimized_2010,clausen_task-optimized_2012,gordon_universal_2007,gordon_optimal_2008,Zwick_ChapGK_2015}
then converges to a sum of delta functions (narrowband filters) centered
at the harmonics of the inverse modulation period, $k\omega_{ctrl}\!=\!\frac{\pi kN}{t}$
with $k=1,2,3,..$ \cite{Alvarez2011,Ajoy2011}. Another suitable
control is CW qubit-driving, which has a single frequency component
($k\!=\!1$). Under such controls, the LM and SM regimes are attained
for $\omega_{ctrl}\tau_{c}\gg1$ and $\omega_{ctrl}\tau_{c}\ll1$,
respectively. The corresponding decay factors, $\mathcal{J}\propto F_{t}(\omega_{ctrl})G(\omega_{ctrl})$
in Eq. (\ref{eq:Dephasing rate xB}), are (see Fig. \ref{fig1_scheme}b)
\begin{eqnarray}
\mathcal{J}^{LM}\!\! & \propto & \!\frac{g^{2}t}{\omega_{ctrl}^{\beta}\tau_{c}^{\beta-1}},\mathcal{\: J}^{SM}\!\!\propto\! g^{2}\tau_{c}t,\label{eq:J}
\end{eqnarray}

\noindent respectively. This reflects the effect of a narrow-band
filter $F_{t}(\omega_{ctrl})$ that may be used to scan the spectral
density $G(\omega)$ \cite{almog_direct_2011,Alvarez2011,Bylander2011},
upon varying the modulating frequency (pulse-rate or Rabi-frequency)
$\omega_{ctrl}$ of the control field, all the way from the frequency-independent
regime $G\propto\tau_{c}$ for $\omega\tau_{c}\ll1$ to the power-law
regime $G\propto\omega^{-\beta}\tau_{c}^{-(\beta-1)}$ for $\omega\tau_{c}\gg1$
(Fig. \ref{fig:FtwGw}a). In the limit of extremely narrow spectral
filters, i.e. $N\!\!\rightarrow\!\infty$, with $\omega_{ctrl}\!=\!\frac{\pi N}{t}$,
we have 
\begin{align}
\left.\frac{\partial\mathcal{J}}{\partial\tau_{c}}\right|_{\omega_{ctrl}\sim\omega_{0}}\!\!\!\!\!\propto\!\!\left.\frac{\partial G}{\partial\tau_{c}}\right|_{\omega_{ctrl}\sim\omega_{0}}\!\!\!\!\!\propto\!\omega_{ctrl}\!-\!\omega_{0};
\end{align}
\begin{equation}
\omega_{0}\!=\!\frac{1}{\tau_{c}(\beta-1)^{\frac{1}{\beta}}}.\label{eq:w0}
\end{equation}
An abrupt change (Fig. \ref{fig:FtwGw}a) is then revealed in the
parametric sensitivity of the attenuation-factor derivative $\frac{\partial\mathcal{J}}{\partial\tau_{c}}$
through its \emph{change of sign}: $\frac{\partial\mathcal{J}}{\partial\tau_{c}}\!\propto\!\!\frac{-(\beta-1)\mathcal{J}^{LM}}{\tau_{c}}\!\!<\!0$
for LM and $\frac{\partial\mathcal{J}}{\partial\tau_{c}}\!\propto\!\frac{\mathcal{J}^{SM}}{\tau_{c}}\!\!>\!0$
for SM, implying that

\begin{equation}
\left.\frac{\partial\mathcal{J}}{\partial\tau_{c}}\right|_{\omega_{ctrl}\tau_{c}(\beta-1)^{\frac{1}{\beta}}\approx1}=0,\label{eq:dJ/dtau}
\end{equation}
at a value dependent on the control (modulating) frequency $\omega_{ctrl}$.
\begin{figure}[t]
\centering{}

\includegraphics[width=1\columnwidth]{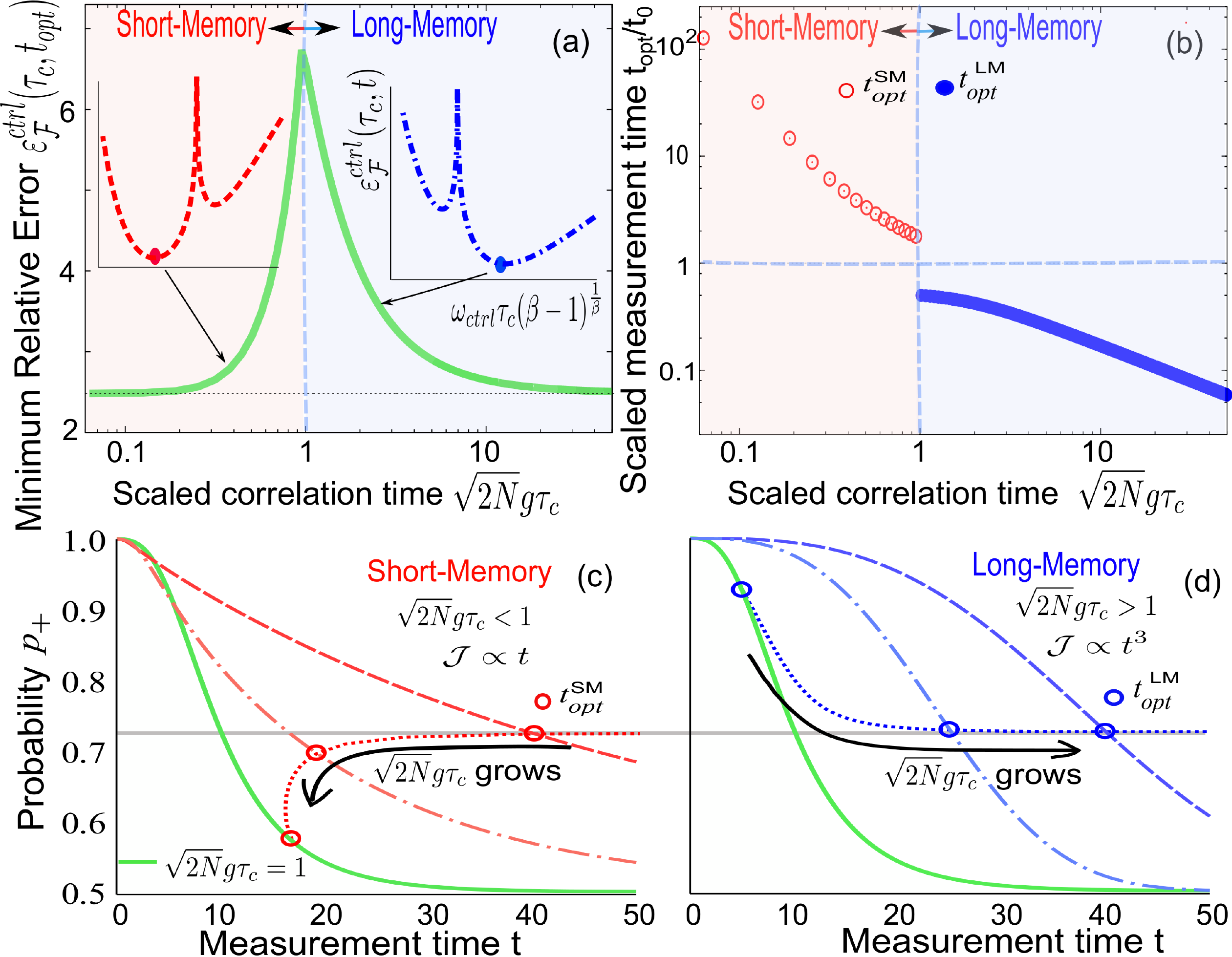}

\protect\caption{\label{fig:e0_p+}Critical transition of the minimal error in environment
memory-time estimation determined by a probe under dynamical control
CW in the narrow filter approximation. The noise spectrum is a Lorentzian
($\beta=2$). (a) The minimum relative error per measurement $\varepsilon_{\mathcal{F}}^{ctrl}(\tau_{c},t_{opt})$
of the environmental correlation time $\tau_{c}$, as a function of
$\sqrt{2N}g\tau_{c}$, obtained by optimizing the control time $t_{opt}$.
It exhibits a critical behavior at $\sqrt{2N}g\tau_{c}\approx1$.
Insets: $\varepsilon_{\mathcal{F}}^{ctrl}(\tau_{c},t=\frac{\pi N}{\omega_{ctrl}})$
for global minima located in the SM (left inset) and LM (right inset)
regimes. The critical point emerges when both local minima are equal
(as in Fig. \ref{fig:FtwGw}b). (b) The optimal scaled measurement
time $\frac{t_{opt}}{t_{0}}$ as a function of $\sqrt{2N}g\tau_{c}$.
(c-d) Probability $p_{+}(t)$ as a function of time. For $\sqrt{2N}g\tau_{c}<1$,
the optimal time $t_{opt}^{SM}$ (red circles) corresponds to a linear
attenuation factor $\mathcal{J}^{SM}\propto t$ (panel (c) and Fig.
(\ref{fig1_scheme})b), while for $\sqrt{2N}g\tau_{c}>1$, the optimal
time $t_{opt}^{LM}$ (blue solid circles) belongs to a regime where
$\mathcal{J}^{LM}\propto t^{\beta+1}$ (panel (d)). At $\sqrt{2N}g\tau_{c}\approx1$
(vertical dashed line) the transition between the two regimes is observed.
The optimal time jumps between $t_{opt}^{LM}$ and $t_{opt}^{SM}$
at the critical point avoiding the time $t_{0}=\pi N\tau_{c}(\beta-1)^{\frac{1}{\beta}}$
(horizontal dashed line) where no-information about $\tau_{c}$ can
be extracted from the probe.}
\end{figure}
\begin{figure}
\includegraphics[width=1\columnwidth]{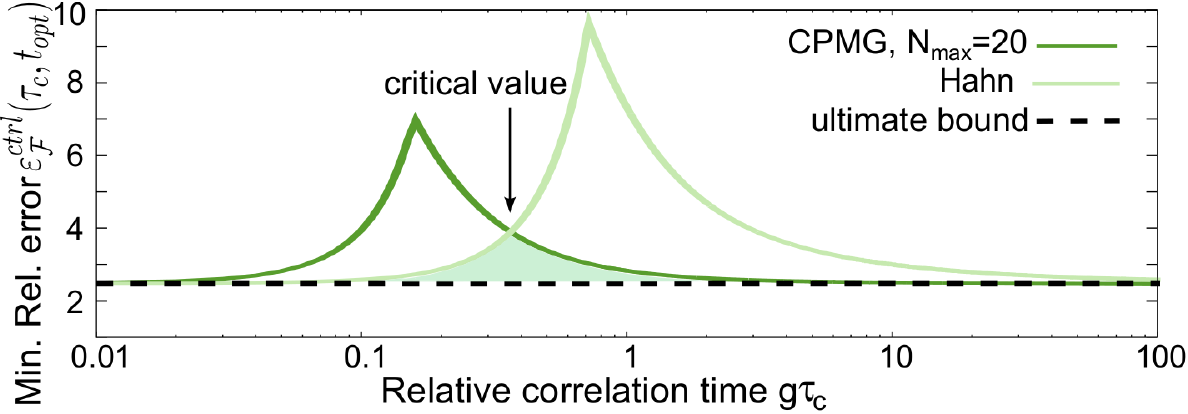}

\protect\caption{\label{fig4}Minimal relative error per measurement in the estimation
of $\tau_{c}$ as a function of $g\tau_{c}$ for a Lorentzian environmental
spectrum under dynamical control. Practical limitations on the number
of pulses of a CPMG sequence, such that $N\le N_{max}$, may prevent
to attainment of the ultimate bound (dashed line) \cite{Zwick_NatCom_2015}.
Then, a sudden change of the dynamical control strategy as a function
of $g\tau_{c}$ may help: For $g\tau_{c}$ lower than the critical
value, the highest precision is achieved by the single-pulse Hahn
echo ($N=1$). However, if $g\tau_{c}$ is larger than the critical
value CPMG sequence with $N=N_{max}$ is optimal. This dynamical control
strategy under practical limitations reduces the minimal error represented
by the shaded area, which is determined by the optimal control on
each side of the intersection (critical value) of the Hahn and the
CPMG curves.}
\end{figure}

The central result of this paper is that Eq. (\ref{eq:dJ/dtau}) signifies
the \emph{vanishing of the quantum Fisher information} (QFI) \cite{Caves_1994_fisher},
which quantifies the attainable amount of information on $\tau_{c}$
that can be extracted from the measured probe-(qubit-) state probabilities
$p_{\pm}$ obeying Eq. (\ref{eq:p+-}). This vanishing becomes apparent
upon considering the expression for QFI \cite{Zwick_NatCom_2015,Zwick_PhysScrip_2015}
\begin{equation}
\mathcal{F_{Q}}(\tau_{c},t)=\frac{e^{-2\mathcal{J}}}{1-e^{-2\mathcal{J}}}\left(\frac{\partial\mathcal{J}}{\partial\tau_{c}}\right)^{2}.\label{eq:QFI_R}
\end{equation}

\noindent Hence, at the value $\omega_{ctrl}\tau_{c}(\beta-1)^{\frac{1}{\beta}}\!\approx\!1$
no information can be extracted, $\mathcal{F_{Q}}^{ctrl}\!=\!0$.
Since the minimum achievable relative error (per measurement) of the
(unbiased) estimation of $\tau_{c}$ is related to the QFI through
the Cramer-Rao bound as 
\begin{equation}
\frac{\delta\tau_{c}}{\tau_{c}}\geq\varepsilon_{\mathcal{F}}(\tau_{c},t)=\frac{1}{\tau_{c}\sqrt{\mathcal{F_{Q}}(\tau_{c},t)}},\label{eq:ultimate_bound}
\end{equation}

\noindent this error \emph{diverges} as $\omega_{ctrl}\rightarrow\omega_{0}$
(Eq. (\ref{eq:w0})): 

\noindent \emph{
\begin{equation}
\varepsilon_{\mathcal{F}}^{ctrl}\left(\tau_{c},t\approx\frac{\pi N}{\omega_{ctrl}}\right)\propto\left|\omega_{ctrl}-\omega_{0}\right|^{-1}.\label{eq:e0_ctrl}
\end{equation}
}

The relative error, $\varepsilon_{\mathcal{F}}^{ctrl}$, thus exhibits
a sharp transition between the LM and SM dynamical regimes (Fig. \ref{fig:FtwGw}b),
allowing their clear distinction. This \textit{critical behavior}
of the relative error provides a signature of\textit{\emph{ }}\textit{the
environmental noise spectral density}\textit{\emph{ through the values
of $\tau_{c}$ and $\beta$,}}\textit{ provided we apply an appropriate
dynamical control} that generates a sufficiently narrow spectral filter,
so as to scan the sign change of $\left.\frac{\partial G}{\partial\tau_{c}}\right|_{\omega\sim\omega_{0}}$
at the critical point (Fig. \ref{fig:FtwGw}). By contrast, such criticality
does not arise under free-induction decay, for which the filter function,
$F_{t}^{free}(\omega)$ is a much broader sinc function centered around
$\omega=0$.

\paragraph{\noindent Critical behavior of the maximal estimation precision of
$\tau_{c}$.--- }

The critical behavior shown above is also manifest, under the same
control on the probe, for the \emph{maximized estimation precision},
\emph{i.e.} the smallest possible minimal relative error in the estimation
of $\tau_{c}$ in Eq. (\ref{eq:ultimate_bound}), 
\begin{equation}
\varepsilon_{\mathcal{F}}^{ctrl}\left(\tau_{c},t_{opt}\right)=\underset{t}{min\,}\varepsilon_{\mathcal{F}}^{ctrl}(\tau_{c},t).\label{eq:min_err}
\end{equation}
The error minimization is the outcome of selecting the \emph{optimal
time} $t_{opt}$ at which the measurement (cf. Eqs. (\ref{eq:ox}),
(\ref{eq:Dephasing rate xB})) is performed on the probe, following
its dephasing under the control we have applied.

\noindent Figure \ref{fig:e0_p+} shows the critical behavior of the
maximum precision per measurement, $\varepsilon_{\mathcal{F}}^{ctrl}(\tau_{c},t_{opt})$
for the \textit{\emph{Lorentzian spectrum}}\foreignlanguage{british}{
($\beta=2$}) following CW control of the qubit-probe as a function
of $\sqrt{2N}g\tau_{c}$. The critical point
\begin{equation}
\sqrt{2N}g\tau_{c}\approx1\label{eq:Ngtau_c}
\end{equation}
separates \emph{two regions characterized by different scaling-laws
of the minimal relative-error as a function of $\sqrt{2N}g\tau_{c}$}
(Fig. \ref{fig:e0_p+}a). These scaling-laws are dictated by the different
\emph{dynamical regimes} for the attenuation (decay) factor shown
in Figs. \ref{fig:e0_p+}c,\ref{fig:e0_p+}d. 

\noindent The optimal probing (measurement) and control time $t_{opt}$
is also shown to undergo a sudden transition at the critical point
(\ref{eq:Ngtau_c}), as shown in Fig. \ref{fig:e0_p+}b. This optimal
time corresponds to the best tradeoff between a signal amplitude contrast,
$\frac{e^{-2\mathcal{J}}}{1-e^{-2\mathcal{J}}}$, and the parametric
sensitivity of the signal attenuation factor, $\left(\frac{\partial\mathcal{J}}{\partial\tau_{c}}\right)^{2}$.
The optimal tradeoff occurs (Fig. \ref{fig:e0_p+}b) at either a long
time, $t_{opt}^{SM}$ (red circle) corresponding to a linear attenuation
factor $\mathcal{J}^{SM}\propto t$ (Fig. \ref{fig:e0_p+}c) or at
a short time, $t_{opt}^{LM}$ (blue circle) corresponding to $\mathcal{J}^{LM}\propto t^{\beta+1}$
(Figs. \ref{fig:e0_p+}d). These optimal control times in the two
regimes are situated on both sides of the critical value (Fig. \ref{fig:e0_p+}b)
\begin{equation}
t_{opt}^{LM}<t_{0}<t_{opt}^{SM},\qquad t_{0}=\frac{\pi N}{\omega_{0}}.\label{eq:t_0-1-1}
\end{equation}

The criticality of the relative error as a function of $\omega_{ctrl}$
described by Eq. (\ref{eq:e0_ctrl}), defines \emph{two} \emph{local,}
unequal minimum values of $\varepsilon_{\mathcal{F}}^{ctrl}$, located
on either side of the critical point (Fig. \ref{fig:FtwGw}b and insets
Fig. \ref{fig:e0_p+}a). What determines the \emph{global} minima
of $\varepsilon_{\mathcal{F}}^{ctrl}$ is the optimal time $t_{opt}$
obtained from Eq. (\ref{eq:min_err}). The critical behavior emerges
when this global minimum \emph{jumps} between the two local minima
as a function of the parameter $\sqrt{2N}g\tau_{c}$ (Fig. \ref{fig:e0_p+}b).
At the critical point, both local minima are equal, as displayed in
Fig. \ref{fig:FtwGw}b.

\emph{Discussion}.--- We have demonstrated a critical behavior of
information (estimation-precision) on the environment fluctuation
(noise) spectrum of generalized \textit{\emph{Ornstein-Uhlenbeck }}process
extracted by a probe subject to appropriate dynamical control as a
function of the environment memory-time $g\tau_{c}$. This finding
applies to any bosonic environment, provided the probe and the environment
are suitably interfaced by a chosen oscillator mode \cite{kofman1994spontaneous,PhysRevE.87.012140}
(cf. discussion following Eq. (\ref{eq:power-law tail noise spectrum})). 

We have shown that similar critical behavior is manifest for the maximal
estimation precision of $\tau_{c}$. At the critical point there is
a massive loss of information on $\tau_{c}$. Near this point, the
optimal time for measuring and controlling the quantum-probe is either
very short, corresponding to little parametric sensitivity, or very
long, corresponding to significant decay of the signal.

The critical behavior of the maximal estimation precision of $\tau_{c}$
has paramount practical implications:

\emph{(i) Complete dynamical behavior characterization}: Rather than
mapping out the long- and short-memory probe dynamics regimes by varying
the probing time, the critical behavior demonstrated here, allows
one to characterize the complete dynamics as consisting of \emph{two}
distinct \emph{dynamical phases} (regimes) according to the maximal
information they yield near the critical point (\ref{eq:Ngtau_c}). 

\emph{(ii) Sudden change of the optimal dynamical control sequence}:
The critical point depends on the control scheme: thus, for CPMG control
\cite{Slichter1990,Carr1954,Meiboom1958} $g\tau_{c}\approx\frac{1}{\sqrt{2N}}$
when probing a\emph{ }\textit{\emph{Ornstein-Uhlenbeck }}process (Lorentzian
spectra). This fact highlights the importance of optimizing the number
of pulses $N$ so as to improve the estimation precision, if $N$
is bounded by $N_{max}$ due to practical limitations. Under these
conditions, the \emph{ultimate bound} on the estimation precision
found in Ref. \cite{Zwick_NatCom_2015}
\begin{equation}
\varepsilon_{\mathcal{F}}(\tau_{c},t)\ge\varepsilon_{0},
\end{equation}
where $\varepsilon_{0}=\frac{\sqrt{1-e^{-2\mathcal{J}_{0}}}}{\mathcal{J}_{0}e^{-\mathcal{J}_{0}}}\approx2.48$,
may not be attained for $g\tau_{c}\approx\frac{1}{\sqrt{2N_{max}}}$
. Then, a sudden change of $N$ should be undertaken as a function
of $g\tau_{c}$ in order to optimize the estimation: For $g\tau_{c}$
lower than a certain critical value shown in Fig. \ref{fig4}, the
best precision is achieved by the single-pulse Hahn echo ($N\!=\!1$).
However, if $g\tau_{c}$ is larger than this critical value, then
the CPMG sequence with $N\!=\! N_{max}$ is optimal. Qualitatively
similar considerations apply for \textit{\emph{generalized Ornstein-Uhlenbeck
}}processes.

To sum up, the critical behavior of the environmental information
revealed here is a fundamental feature that facilitates the attainment
of the highest estimation precision of the environment memory-time.
It represents an alternative characterization of the probe-qubit dynamics
under suitable control or observation that leads to a phase transition
on the dynamical behavior \cite{Alvarez2006,Danieli2007,Rotter2009,Garrahan2010,Lesanovsky2013,'Alvarez2015}.
Intriguingly, the \emph{absence of information} has been shown to
provide a distinctive signature of the environmental noise estimation.

Such information may be useful e.g., for studying molecular diffusion
at the nanoscale and thereby characterizing biological systems \cite{Alvarez2013,Alvarez2013_JMR,Zwick_NatCom_2015,Shemesh2015}
or chemical-shift effects \cite{Smith2012}; charge diffusion in conducting
crystals \cite{Feintuch2004} or spin diffusion in complex spin-networks
\cite{Suter1985,Alvarez2011,'Alvarez2015,Alvarez2013a,alvarez_controlling_2012}.
Knowledge of the memory time may also be important for studying fundamental
effects, such as quantum phase transitions in a spin environment \cite{Haikka2012,Gessner2014}
or nonlocal correlations within a composite environment \cite{Smirne2013,Laine2012,Liu2013,Wissmann2014}.
\begin{acknowledgments}
G.A.A. acknowledges the support of the European Commission under the
Marie Curie Intra-European Fellowship for Career Development grant
no. PIEF-GA-2012-328605. G.K. acknowledges the ISF support under the
Bikura (Prime) grant.\end{acknowledgments}

\end{document}